\documentclass[UTF-8,preprint,showpacs,preprintnumbers,superscriptaddress]{revtex4-1}
\usepackage{multirow}
\usepackage{diagbox}
\usepackage{booktabs}
\usepackage{array}
\usepackage{natbib}
\usepackage{amsmath,amssymb,graphicx,bm,subfigure,float,siunitx,color}
\usepackage{bm,subfigure,float,siunitx,graphicx}
\usepackage[utf8]{inputenc}
\usepackage[figuresright]{rotating}
\usepackage{color}
\usepackage{epstopdf}
\usepackage{mathrsfs}
\begin{document}

\title{Kiselev Black holes in quantum fluctuation modified gravity}

\author{Yaobin Hua}
\affiliation{College of Physics Science and Technology, Hebei University, Baoding 071002, China}
\author{Rong-Jia Yang \footnote{Corresponding author}}
\email{yangrongjia@tsinghua.org.cn}
\affiliation{College of Physics Science and Technology, Hebei University, Baoding 071002, China}
\affiliation{Hebei Key Lab of Optic-Electronic Information and Materials, Hebei University, Baoding 071002, China}
\affiliation{National-Local Joint Engineering Laboratory of New Energy Photoelectric Devices, Hebei University, Baoding 071002, China}
\affiliation{Key Laboratory of High-pricision Computation and Application of Quantum Field Theory of Hebei Province, Hebei University, Baoding 071002, China}

\begin{abstract}
We obtain a new general solution for the gravitational field equations in quantum fluctuation modified gravity, which reduces to different classes of black holes surrounded by fluids, by taking some specific values of the parameter of the equation of state. We discuss the strong energy condition in a general way and also for some special cases of different fluids. In addition, the Hawking temperature associated to the horizons of solutions and constraints on the parameter characterizing the fluctuation of metric are taken into account in our analysis.
\end{abstract}


\maketitle

\section{Introduction}
Numerous observations have confirmed that the universe is accelerating its expansion. To explain this phenomenon, modified gravity theories, such as $f(R)$ theories, Gauss-Bonnet gravity, and $f(R, T)$ theory, are popular attempts. Recently, quantum fluctuation modified gravity (QFMG) was proposed from Heisenberg's nonperturbative quantization \citep{Dzhunushaliev:2013nea,Dzhunushaliev:2015mva,Dzhunushaliev:2015hoa}: if the metric can be decomposed into the sum of the classical part and the quantum fluctuation part, then the corresponding quantum Einstein gravity engenders at the classical level modified gravity models with a non-minimal coupling between geometry and matter \cite{Yang:2015jla}. This idea has been implemented in some special models, see for example \cite{Yang:2015jla,Liu:2016qfx,Bernardo:2021ynf,Chen:2021oal,Lima:2023qdv}. Black hole (BH) solutions in QFMG was discussed in \cite{Yang:2020lxv}. Friedmann-Lemaitre-Robertson-Walker cosmology in the framework of QFMG was considered in  \cite{Yang:2015jla,Liu:2016qfx,Bernardo:2021ynf,Haghani:2021iqe,Chen:2021oal}. Recently, baryogenesis in QFMG was investigated in \cite{Yang:2024gnf}.

BH solutions are important research subjects in gravitational theories, serving as a crucial testing ground for our understanding of modified gravity and general relativity (GR). In QFMG, there exists a coupling between matter and geometry. It is expected differences between solutions in QFMG and GR in the presence of the matter-geometry coupling. Such differences may become more prominent for high densities, in this way, it is natural to test the effects imposed by QFMG in the scale of compact objects. The effects of matter-geometry on the structure of BHs has been discussed in \cite{Yang:2020lxv}. Here we will focus on seeking Kiselev BHs in QFMG. Kiselev suggested a relation connecting energy density and pressure \cite{Kiselev:2002dx}: the components of the energy-momentum tensor is associated to an anisotropic fluid, and one can obtain the barotropic equation of state (EoS) by taking the isotropic average over the angles. Some particular choices of the parameter of EoS can be assumed in this formulation and some of these values reproduce the accelerating pattern in the cosmological context \cite{Kiselev:2002dx}. Recently various Kiselev BHs have been proposed and widely studied \cite{Santos:2023fgd,Ghosh:2015ovj,Xu:2016jod,Toshmatov:2015npp,Ghosh:2023nkr,Singh:2023zmy,Hamil:2024ppj,Giambo:2023zmy,Qu:2023rsv,Heydarzade:2023dof,Benali:2024clh,
Sood:2024ufi,Visser:2019brz,Bezerra:2022srj,Lungu:2024iob,Chaudhary:2024yod,Gogoi:2024ypn,Hamil:2022pzr,Sheikhahmadi:2023jpb,Zahid:2023csk,
Hamil:2024njs,Ghosh:2017cuq,Heydarzade:2017wxu,Sakti:2019krw,Visser:2019brz}

Here we will try to obtain nontrivial BH solutions for the field equations
in the context of QFMG where the BH is surrounded by the fluid discussed by
Kiselev \cite{Kiselev:2002dx}. We will discuss some cases associated to the solution obtained by taking into account the appropriate values of the EoS of fluid. We will also investigate the energy conditions fulfilled by the fluid, the BH thermodynamics and their constraints on the parameters.

The paper is organized as follows. In the next section, we will briefly review the QFMG proposed in \cite{Yang:2015jla}. In Sec. III, we will present a new Kiselev BH solution. In Sec. IV, we will discuss the strong energy condition in a general way. In Sec. V, we will investigate the thermodynamic quantities associated to the BH solutions and constraints on the parameter characterizing the fluctuation of metric. Finally, we will briefly summarize and discuss our results in section VI.

\section{Quantum fluctuation modified gravity}
For Heisenberg's nonperturbative quantization, the metric operator is split into a sum of an classical part $g_{\mu\nu}$ and a quantum fluctuating part $\delta\widehat{ g}_{\mu\nu}$ \citep{Dzhunushaliev:2013nea}, $\hat{g}_{\mu\nu}=g_{\mu\nu}+\delta \widehat{g}_{\mu\nu}$. In \cite{Wetterich:2016vxu}, $\delta \hat{g}^{\mu\nu}$ is decomposed into a ``physical metric" and a ``gauge part". One can eliminate the gauge part by a suitable gauge transformation. In general, one has $\langle\delta \hat{g}_{\mu\nu}\rangle\neq 0$ for any quantum non vacuum state. At first-order, the quantum Einstein-Hilbert Lagrangian can be expanded as
\begin{eqnarray}
\label{lag0}
L_{\hat{g}}(\hat{g})=L_{\hat{g}}(g+\delta\hat{g})\approx L_{g}(g)+\frac{\delta L_{g}}{\delta g^{\mu\nu}}\delta \widehat{g}^{\mu\nu}.
\end{eqnarray}
Here we take unit $c=\hbar=1$ and $k^2=8\pi G$. Because $\left\langle \frac{\delta L_{g}}{\delta g^{\mu\nu}}\delta \widehat{g}^{\mu\nu}\right\rangle=\frac{\delta L_{g}}{\delta g^{\mu\nu}}\langle\delta \widehat{g}^{\mu\nu}\rangle=\sqrt{-g}G_{\mu\nu}\langle\delta \widehat{g}^{\mu\nu}\rangle$ \cite{Dzhunushaliev:2013nea}, the expectation value of Lagrangian (\ref{lag0}) is given by
\begin{eqnarray}
\label{lag00}
\langle L_{\hat{g}}\rangle\approx \frac{1}{2k^2}\sqrt{-g} \Big[R+G_{\mu\nu}\langle\delta \hat{g}^{\mu\nu}\rangle \Big],
\end{eqnarray}
Similarly the quantum Lagrange density $L^{\hat{g}}_{\rm m}$ for matter can be expanded as
\begin{eqnarray}
\label{lagm}
L^{\hat{g}}_{\rm m}(g+\delta\hat{g})\approx \sqrt{-g}L_{\rm m}(g)+\frac{\delta \sqrt{-g}L_{\rm m}}{\delta g^{\mu\nu}}\delta \widehat{g}^{\mu\nu}.
\end{eqnarray}
The expectation value of the quantum Lagrange density (\ref{lagm}) takes the form \citep{Dzhunushaliev:2013nea}
\begin{eqnarray}
\label{lagm0}
\langle L^{\hat{g}}_{\rm m}(g+\delta\hat{g}) \rangle \approx \sqrt{-g}\bigg[L_{\rm m}-\frac{1}{2}T_{\mu\nu}\langle\delta \hat{g}^{\mu\nu}\rangle\bigg],
\end{eqnarray}
were $T_{\mu\nu}=-2\delta(\sqrt{-g}L_{\rm m})/(\sqrt{-g}\delta g^{\mu\nu})$ is the stress-energy tensor. Then we have the modified Lagrangian density as
\begin{eqnarray}
\label{lag}
L=\frac{1}{2k^2}\sqrt{-g} \bigg[R+G_{\mu\nu}\langle\delta \hat{g}^{\mu\nu}\rangle \bigg]+\sqrt{-g}\bigg[L_{\rm m}-\frac{1}{2}T_{\mu\nu}\langle\delta \hat{g}^{\mu\nu}\rangle\bigg].
\end{eqnarray}
For a simple case, $\langle\delta \hat{g}^{\mu\nu}\rangle=\alpha g^{\mu\nu}$ with $\alpha$ a constant indicating the magnitude of quantum fluctuations, which fulfills the two conditions suggested in \cite{Wetterich:2016vxu}, $\nabla_\alpha\langle\delta \hat{g}^{\mu\nu}\rangle=0$ and $\langle\delta \hat{g}^{\mu\nu}\rangle=\langle\delta \hat{g}^{\nu\mu}\rangle$, the Lagrangian density (\ref{lag}) becomes \cite{Yang:2015jla},
\begin{eqnarray}
\label{act}
L=L_{\rm mg}+L_{\rm mm}=\frac{1}{2k^2}\sqrt{-g}(1-\alpha)R+\sqrt{-g}\bigg[L_{\rm m}-\frac{1}{2}\alpha T\bigg],
\end{eqnarray}
were $T=g_{\mu\nu}T^{\mu\nu}$ is the trace of the stress-energy tensor. We consider the case $|\alpha|<1$, meaning that the quantum fluctuation is smaller than the classical part of the metric operator. Assuming $\delta g_{\mu\nu}=0$ on the boundary and varying the Lagrangian density (\ref{lag}) with respect to $g^{\mu\nu}$, we derive the gravitational field equations \cite{Yang:2015jla}
\begin{eqnarray}
\label{mot}
G_{\mu\nu}\equiv R_{\mu\nu}-\frac{1}{2}g_{\mu\nu}R=\frac{2k^2}{1-\alpha}\left[\frac{1}{2}(1+\alpha)T_{\mu\nu}-\frac{1}{4}\alpha g_{\mu\nu}T+\frac{1}{2}\alpha\theta_{\mu\nu}\right],
\end{eqnarray}
where $\theta_{\mu\nu}=g^{\alpha\beta}\delta T_{\alpha\beta}/\delta g^{\mu\nu}$ and $\theta=g_{\mu\nu}\theta^{\mu\nu}$. With $R=-k^2[T+\alpha\theta/(1-\alpha)]$, we rewritten the gravitational field equations (\ref{mot}) as \cite{Yang:2015jla}
\begin{eqnarray}
\label{mot1}
R_{\mu\nu}=\frac{2k^2}{1-\alpha}\left[\frac{1}{2}(1+\alpha)T_{\mu\nu}-\frac{1}{4} g_{\mu\nu}T+\frac{1}{2}\alpha\theta_{\mu\nu}-\frac{1}{4}\alpha g_{\mu\nu}\theta\right].
\end{eqnarray}
The Lagrangian density (\ref{lag}) indicates an possible microscopic quantum description of the matter creation processes in $f(R, T)$ or $f(R, L_{\rm{m}})$ gravity \cite{Liu:2016qfx,Yang:2020lxv}. Such a description may shed light on the physical mechanisms leading to particle generation via matter-geometry coupling.

\section{New Kiselev black hole solution}
For a spherically symmetric and static space-time, the line element can be written as
\begin{eqnarray}
\label{lineelement}
ds^{2}=B(r)dt^{2}-A(r)dr^{2}-r^{2}\Big(d\theta^{2}+\sin^{2}\theta d\phi^{2}\Big),
\end{eqnarray}
where $A(r)$ and $B(r)$ are analytical functions of coordinate $r$. The components of energy-momentum tensor in Kiselev BH are assumed to take the form \cite{Kiselev:2002dx}
\begin{eqnarray}
	&& T^{t}_{t}=T^{r}_{r}=\rho(r),                                        \label{T12} \\
	&& T^{\theta}_{\theta}=T^{\phi}_{\phi}=-\frac{1}{2}\rho(3\omega+1),    \label{T34}
\end{eqnarray}
where $\rho$ is the energy density, $\omega=p/\rho$ is the parameter of EoS obtained by taking the isotropic average
over the angles in the place of Eqs. (\ref{T12}) and (\ref{T34}) \cite{Kiselev:2002dx}, and $p$ is the pressure. The the components of energy-momentum tensor
effectively connected to an anisotropic fluid in Kiselev BH are represented by
\begin{eqnarray}
	\label{TTT}
	T^{\mu}_{\nu}={\rm{diag}}(\rho,-p_{r},-p_{t},-p_{t}),
\end{eqnarray}
where $p_{t}$, $\rho$ and $p_{r}$ are the transverse pressure, the energy density, and the radial pressure of the fluid, respectively. Comparing with Eqs. \eqref{T12} and \eqref{TTT}, we have $p_{r}=-\rho$, $p_{t}=\frac{1}{2}\rho(3\omega+1)$. The Lagrangian density for matter associated to the anisotropic fluid is given by $L_{m}=(-1/3)(p_{r}+2p_{t})$ from which we can derive $\theta_{\mu\nu}=-2T_{\mu\nu}-\frac{1}{3}(p_{r}+2p_{t})g_{\mu\nu}$. In the context of the Kiselev solutions of black holes, these results demand additivity and linearity between the metric components.

Inserting the energy-momentum tensor \eqref{TTT} into the field equation \eqref{mot}, yields
\begin{eqnarray}
    &&G_{t}^{t}=G_{r}^{r}=H(\rho),                          \label{GGH}  \\
    &&G_{\theta}^{\theta}=G_{\phi}^{\phi}=F(\rho),          \label{GGF}
\end{eqnarray}
where
\begin{eqnarray}
	&&H(\rho)=\frac{2k^2}{1-\alpha}\left[\frac{1}{2}\rho(1+\alpha)-\frac{1}{4}\alpha\rho(1-3\omega)-\frac{1}{2}\alpha\rho(2+\omega)\right],                       \label{Hp}          \\
	&&F(\rho)=\frac{2k^2}{1-\alpha}\left[-\frac{1}{4}\rho(1+\alpha)(3\omega+1)-\frac{1}{4}\alpha\rho (1-3\omega)+\frac{1}{2}\alpha\rho(2\omega+1)\right].         \label{Fp} 	
\end{eqnarray}

For black holes surrounded by a fluid whose components of the energy-momentum tensor are given by
Eqs. \eqref{T12} and \eqref{T34}, Eqs. \eqref{GGH} and \eqref{GGF} form the independent set of field equations give the relation
\begin{eqnarray}
	\label{solveA}
	B(r)\frac{dA(r)}{dr}+A(r)\frac{dB(r)}{dr}=0.
\end{eqnarray}
As a result, we have $A(r)=1/B(r)$. Substituting this result in the Einstein tensor, we derive

\begin{eqnarray}
	&&G_{t}^{t}=G_{r}^{r}=-\frac{1}{r}\frac{dB(r)}{dr}-\frac{B(r)}{r^{2}}+\frac{1}{r^{2}}=H(\rho).                          \label{newGGH}  \\
	&&G_{\theta}^{\theta}=G_{\phi}^{\phi}=-\frac{1}{2}\frac{d^{2}B(r)}{dr^{2}}-\frac{1}{r}\frac{dB(r)}{dr}=F(\rho).          \label{newGGF}
\end{eqnarray}

There are some conditions, such as additivity and linearity, restricts the form of functions $H$ and $F$  \cite{Kiselev:2002dx}. If assuming $H=\lambda F$, where $\lambda$ is an constant, we have
\begin{eqnarray}
	\label{solveM}
	\lambda=\frac{H}{G}=\frac{2-3\alpha+\omega\alpha}{4\omega\alpha-3\omega-1}.
\end{eqnarray}
Combining this equation with Eqs. (\ref{GGH}) and (\ref{GGF}), gives
\begin{eqnarray}
\label{integrate1} -\frac{1}{r}\frac{dB(r)}{dr}-\frac{B(r)}{r^{2}}+\frac{1}{r^{2}}=
\frac{2-3\alpha+\omega\alpha}{4\omega\alpha-3\omega-1}\left[-\frac{1}{2}\frac{d^{2}B(r)}{dr^{2}}-\frac{1}{r}\frac{dB(r)}{dr}\right].
\end{eqnarray}
Eq. (\ref{integrate1}) can be rewritten as
\begin{eqnarray}
	\label{integrate2}
	\frac{1}{r^{2}}\left[\frac{d}{dr}\left(rB(r)\right)-1\right]=
-\frac{1}{2r}\frac{2-3\alpha+\omega\alpha}{1-4\omega\alpha+3\omega}\frac{d}{dr}\left[\frac{d}{dr}\left(rB(r)\right)-1\right].
\end{eqnarray}
After integrating, we get
\begin{eqnarray}
	\label{integrate3}
	\frac{d}{dr}\left[rB(r)\right]-1=a r^{\frac{-2(1+3\omega-4\omega\alpha)}{2-3\alpha+\omega\alpha}},
\end{eqnarray}
where $a$ is an integration constant. Integrating again, we have the solution as
\begin{eqnarray}
	\label{eqB}	
	B(r)=1+\frac{C}{r}+Dr^{-\frac{2(1+3\omega-4\omega\alpha)}{2-3\alpha+\omega\alpha}},
\end{eqnarray}
where $C$ and $D$ are integration constants. If there is no fluid surrounding the BH, solution \eqref{eqB} will reduce to the Schwarzschild BH, so we get $C=-2M$, where $M$ is total mass of the BH. Compared with the Kiselev BH in GR, due to the emergence of parameter $\alpha$ derived from metric quantum fluctuations, solution \eqref{eqB} is more complex, interesting, and has a richer structure. When $\alpha\rightarrow 0$, the solution \eqref{eqB} reduces to the Kiselev BH in GR which admits at most two horizons in the interval $0<\omega\le1$. Under what conditions does solution \eqref{eqB} also exhibit similar properties?  Since $M>0$, $-1<\alpha<1$, $\omega<1$, and $\alpha\neq2/(3-\omega)$, (i) if the BH \eqref{eqB} has two horizons, the parameters should fulfil the following additional conditions: $D>0$, $\beta=2(1+3\omega-4\omega\alpha)/(2-3\alpha+\omega\alpha)>1$, $D<(2M/\beta)[2M(\beta-1)/\beta]^{\beta-1}$; (ii) if Eq. \eqref{eqB} represents an extremal BH, the parameters should also fulfil these conditions:
$\beta>1$, $D>0$, and $D=(2M/\beta)[2M(\beta-1)/\beta]^{\beta-1}$, or $\beta<1$ and $D>0$; (iii) if Eq. \eqref{eqB} is a naked singularity, the parameters should fulfil these additional conditions: $\beta>1$ and $D>(2M/\beta)[2M(\beta-1)/\beta]^{\beta-1}$. For $0<\omega\le1$, the Kiselev BH in GR is asymptotically flat. The parameter $\alpha$ in the metric \eqref{eqB} may change this characteristic. However, If the parameters $\omega$ and $\alpha$ satisfy the condition: $0<2(1+3\omega-4\omega\alpha)/(2-3\alpha+\omega\alpha)\le 1$, the spacetime \eqref{eqB} will also be asymptotically flat.

Substituting (\ref{eqB}) into the field equation (\ref{newGGH}), we get the energy density of fluid
\begin{eqnarray}
	\label{eq rho rho}	
	\rho=Kr^{-\frac{{6(1+\omega)(1-\alpha)}}{{2-3\alpha+\omega\alpha}}},
\end{eqnarray}	
where
\begin{eqnarray}
	\label{eq rho V}	
	K=\frac{1-\alpha}{k^{2}}\frac{6D(\alpha+2\omega-3\omega\alpha)}{(2-3\alpha+\omega\alpha)^{2}}.
\end{eqnarray}	
Solution \eqref{eqB} is dependent on parameters $\alpha$ and $\omega$. For different values of $\alpha$ and $\omega$, we have different solutions for BH.

\section{Energy conditions}
It is well-known that exotic matter violates certain energy conditions of the energy-momentum tensor. Regarding the energy conditions for the anisotropic fluid, the components of energy-momentum for anisotropic fluids must satisfy some requirements in order to represent the true distribution of matter. There are usually the following energy conditions \cite{Martin-Moruno:2017exc,Barcelo:2002bv,Martin-Moruno:2013wfa,Martin-Moruno:2013sfa}. The weak energy condition (WEC) ($\rho\ge0$) measured by any timelike observer ensures that the energy density is always non-negative. The null energy condition (NEC) ($\rho+p\ge0$) ensures that the energy density and pressure along any null (lightlike) direction contribute positively. The dominant energy condition (DEC) ($\rho-p\ge0$) requires that the energy density exceeds or equals the pressure in magnitude. The strong energy condition (SEC) ($\rho+3p\ge0$) has a deeper connection to the dynamics of spacetime curvature and the expansion or contraction of the universe. Among these conditions, DEC is the strongest limitation if $p\ge0$, while SEC is strongest limitation if $p\le0$. Here we focus on the SEC. Considering the SEC for anisotropic fluids, the expression is given by the following equations

\begin{eqnarray}
	 \rho+p_{n}\geq0, ~~~~\rho+\sum_{n}p_{n}\geq0,       \label{oldSEC2}	
\end{eqnarray}	
where $n=1, 2, 3...$. Combining Eqs. (\ref{eq rho rho}), (\ref{T12}), (\ref{T34}) and (\ref{TTT}), we derive the radial and tangential pressure, respectively, as
\begin{eqnarray}
	&& p_{r}=-\rho=-Kr^{-\frac{{6(1+\omega)(1-\alpha)}}{{2-3\alpha+\omega\alpha}}},                  \label{oldSEC3}  \\
	&& p_{t}=\frac{1}{2}(3\omega+1)Kr^{-\frac{{6(1+\omega)(1-\alpha)}}{{2-3\alpha+\omega\alpha}}},	\label{oldSEC4}
\end{eqnarray}	
from which we can easily get
\begin{eqnarray}
	&& \rho+p_{r}=0,                          \label{oldSEC5}	   \\
	&& \rho+p_{t}=\frac{3}{2}(\omega+1) Kr^{-\frac{{6(1+\omega)(1-\alpha)}}{{2-3\alpha+\omega\alpha}}},    \label{oldSEC6}	   \\
	&& \rho+p_{r}+2p_{t}=(3\omega+1) Kr^{-\frac{{6(1+\omega)(1-\alpha)}}{{2-3\alpha+\omega\alpha}}}.    \label{oldSEC7}
\end{eqnarray}	
According to Eqs. (\ref{oldSEC5}), (\ref{oldSEC6}), and (\ref{oldSEC7}), the conditions satisfied by the SEC are given by
\begin{eqnarray}
	&&\frac{(3\omega+1)(\alpha+2\omega-3\omega\alpha)(1-\alpha)D}{(2-3\alpha+\omega\alpha)^{2}}\geq0    \label{sec1},  \\
	&& \frac{(\omega+1)(\alpha+2\omega-3\omega\alpha)(1-\alpha)D}{(2-3\alpha+\omega\alpha)^{2}}\geq0	\label{sec2}.
\end{eqnarray}	

\begin{figure}[h]
    \centering
    \begin{minipage}[b]{0.40\textwidth}
        \centering
        \includegraphics[width=\textwidth]{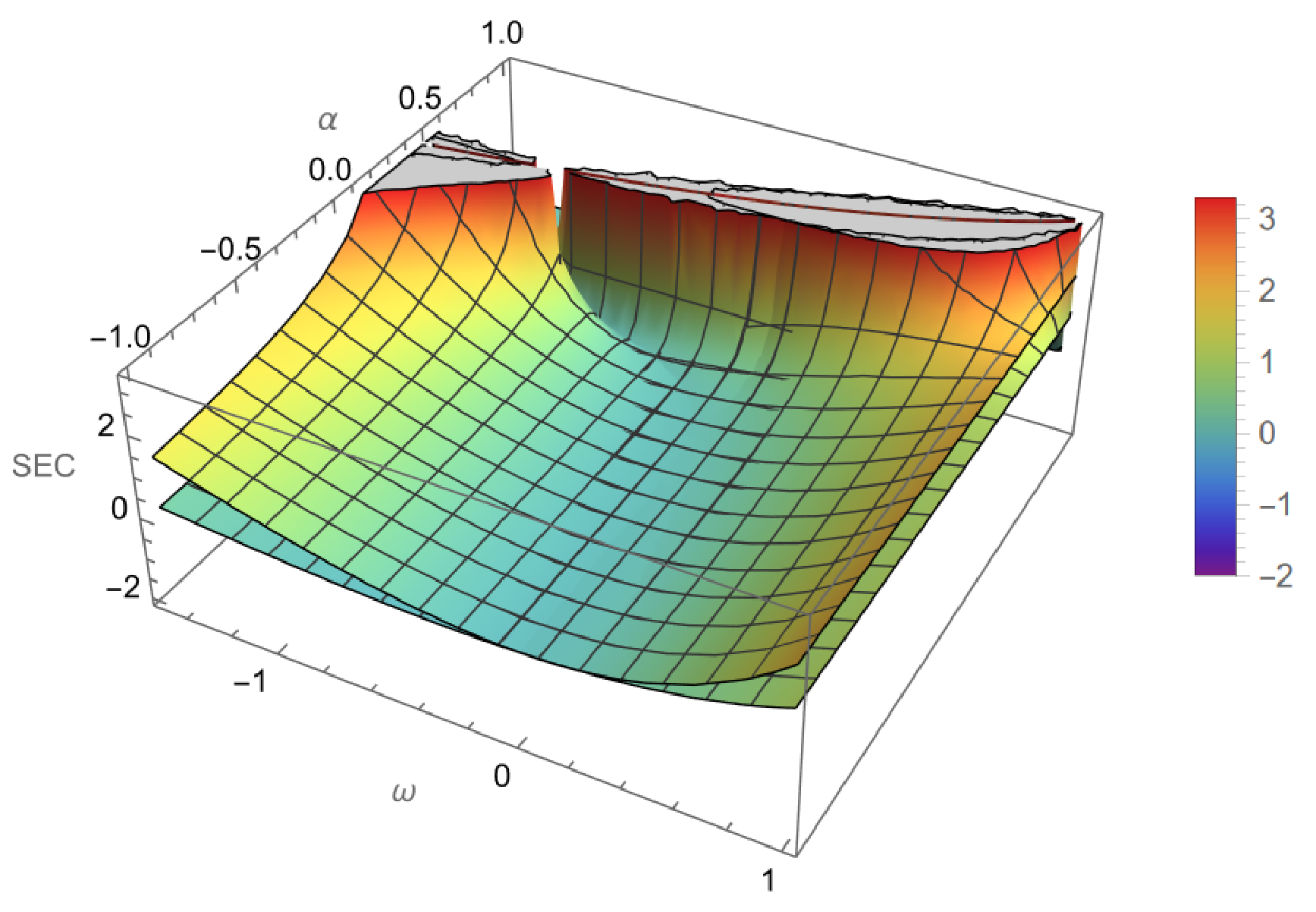}
    \end{minipage}
     \hspace{3em}
    \begin{minipage}[b]{0.40\textwidth}
        \centering
        \includegraphics[width=\textwidth]{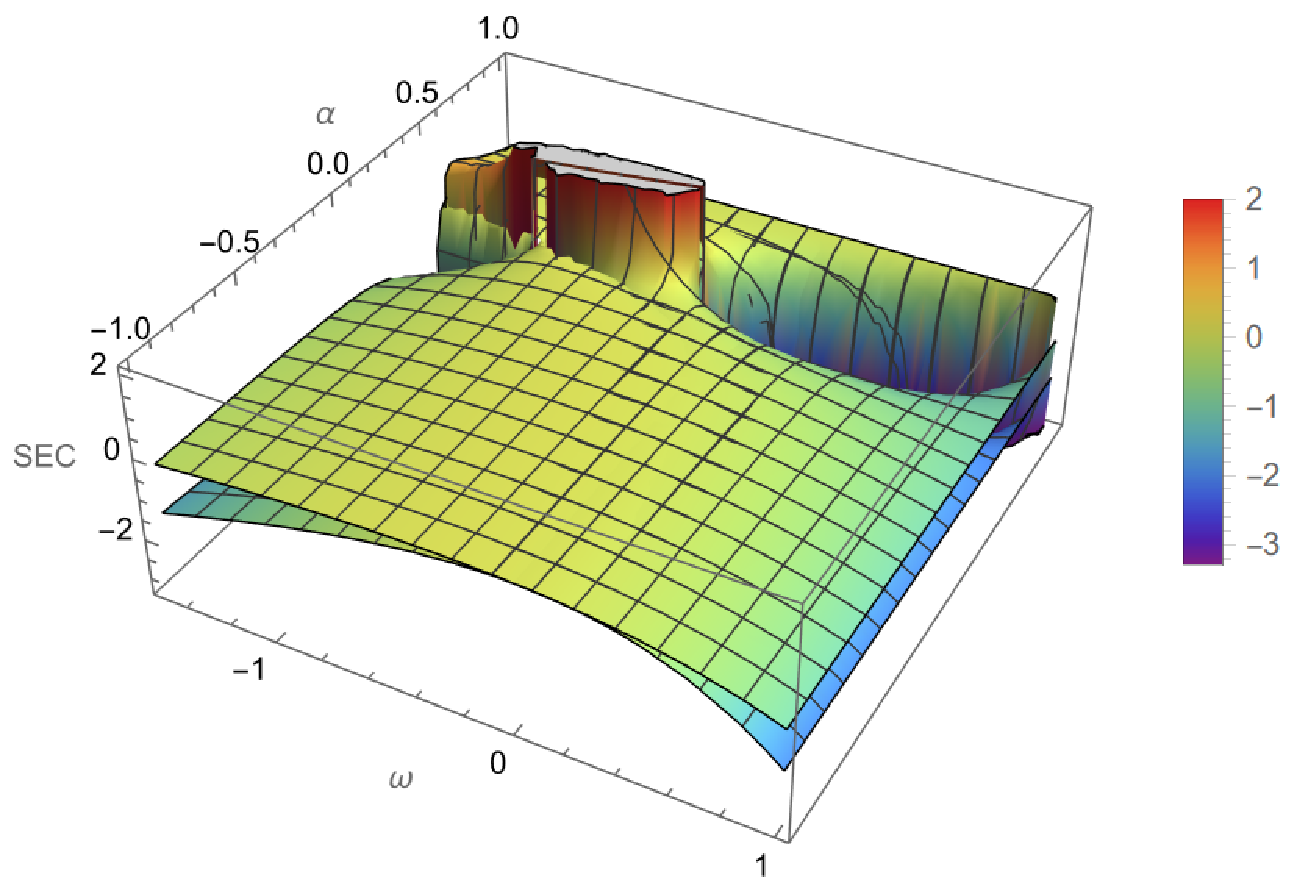}
    \end{minipage}
\caption{Conditions (\ref{sec1}) and (\ref{sec2}) is plotted as a function of $\omega$ and $\alpha$ with $D=1$ (left) or $D=-1$ (right).}
\label{SEC}
\end{figure}

In Fig. \ref{SEC}, we visualize the possible solutions that satisfy the above equations connecting the parameters $\omega$, $\alpha$, and SEC. We plot the left-hand sides (LHSs) of Eqs. (\ref{sec1}) and (\ref{sec2})  with $D=1$ or $D=-1$.  In these plots, positive values of the
independent variable (vertical axis) correspond to the region where the SEC holds true.

\section{Thermodynamic quantities and constraints}
The position where the metric function $B(r)$ equals zero is denoted as $r_{\rm{h}}$ and $B(r_{\rm{h}})=0$ represents the event horizon of the metric function. From this definition, the mass of BH can be written in terms of $r_{\rm{h}}$ as follows
\begin{eqnarray}
	\label{eqM}		
M(r_{\rm{h}})=\frac{1}{2}r_{\rm{h}}\left[1+Dr_{\rm{h}}^{-\frac{2\left(1+3\omega-4\omega\alpha\right)}{2-3\alpha+\omega\alpha}}\right].
\end{eqnarray}	
This equation shows a general relation connecting the parameters $\omega$, $\alpha$ and $M$. The surface gravity of BH surround by a fluid is given by
\begin{eqnarray}
 	\label{eq kappa}
 	\kappa=\frac{1}{2}{\frac{dB(r)}{dr}\bigg|}_{r=r_{\rm{h}}},
\end{eqnarray}
from which we have
\begin{eqnarray}
	\label{eq kappa finally}
	\kappa=\frac{1}{2 r_{\rm{h}}}-\frac{3(\alpha+2\omega-3\omega\alpha)Dr_{\rm{h}}^{-\frac{2(1+3\omega-4\omega\alpha)}{2-3\alpha+\omega\alpha}}}{2 r_{\rm{h}}(2-3\alpha+\omega\alpha)},
\end{eqnarray}
This quantity, in terms of the parameter of EoS $\omega$ and of the parameter characterizing the quantum fluctuation of metric $\alpha$, gives the Hawking temperature as $T= \kappa/(2\pi)$, which follows
\begin{eqnarray}
	\label{eqT}		
	T_{\rm{BH}}=\frac{\kappa}{2\pi}=\frac{1}{4\pi r_{\rm{h}}}\left[1-\frac{3(\alpha+2\omega-3\omega\alpha)Dr_{\rm{h}}^{-\frac{2(1+3\omega-4\omega\alpha)}{2-3\alpha+\omega\alpha}}}{2-3\alpha+\omega\alpha}\right] \equiv \frac{1}{4\pi r_{\rm{h}}}\left[1-\mathcal{M}'(r)\right],
\end{eqnarray}	
From this expression, we find that the Hawking temperature of a BH surrounded by a fluid in QFMG has an additional structure coming from the dependence on the parameter $\alpha$. Here we focus on the canonical type of horizon \cite{Visser:1992qh,Vertogradov:2024fbd}: $1-\mathcal{M}'(r)\geq0$, meaning that the Hawking temperature is always nonnegative. In the following, we will analyze the influence of parameter $\alpha$ on this result by taking particular choices of the parameter $\omega$.

\subsection{Black hole surrounded by a dust field}
When a BH surrounded by a dust field, $\omega=0$, Eq. (\ref{eqB}) reduces to the following form
\begin{eqnarray}
	\label{eqB1}	
	B(r)=1-\frac{2M}{r}+Dr^{-\frac{2}{2-3\alpha}},
\end{eqnarray}
which implies that $\alpha\neq 2/3$, otherwise the exponent will become singular. The presence of $\alpha$ means that this solution is not equal to the metric of a BH surrounded by a dust field in GR \cite{Kiselev:2002dx}. For the case of $\alpha\rightarrow0$, Eq. (\ref{eqB1}) reduces to the metric of a Schwarzschild BH with an effective mass $2M_{\rm{eff}}=2M-D$.

The SEC condition, represented by Eqs. (\ref{sec1}) and (\ref{sec2}) which are equivalent for this case ($\omega =0$), is given by
\begin{eqnarray}
	\label{sec01}	
	\frac{\alpha(1-\alpha)D}{(2-3\alpha)^{2}}\geq0,
\end{eqnarray}	
from which the parameter $\alpha$ is constrained as $0\leq\alpha<1$ with $\alpha\neq2/3$ for $D>0$, while it is constrained as $-1<\alpha\leq0$ for $D<0$.  In this case, the SEC is satisfied in GR ($\alpha=0$). In QFMD, the SEC is satisfied or not, depending on the values of $\alpha$.

\begin{figure}[h]
    \centering
    \begin{minipage}[b]{0.40\textwidth}
        \centering
        \includegraphics[width=\textwidth]{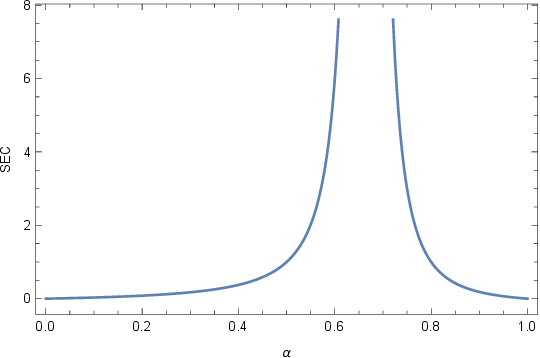}
    \end{minipage}
     \hspace{3em}
    \begin{minipage}[b]{0.40\textwidth}
        \centering
        \includegraphics[width=\textwidth]{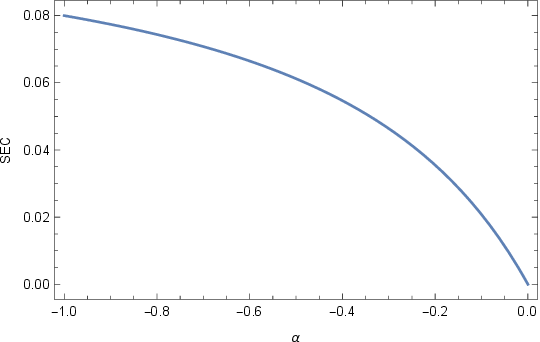}
    \end{minipage}
\caption{Condition (\ref{sec01}) is plotted as a function of $\alpha$ with $\omega=0$, $D=1$ (left) or $D=-1$ (right).}
\label{dust1}
\end{figure}

According the condition imposed by the inequality \eqref{sec01}, we plot the region where the SEC is satisfied in Fig. \ref{dust1} with $D=1$ or $D=-1$. In these plots, positive values of the expression correspond to the region where the SEC holds true.

\begin{figure}[h]
    \centering
    \begin{minipage}[b]{0.40\textwidth}
        \centering
        \includegraphics[width=\textwidth]{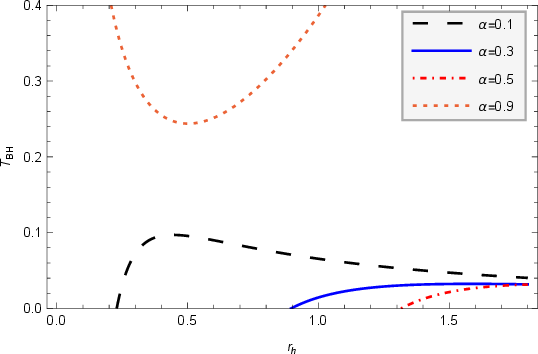}
    \end{minipage}
     \hspace{3em}
    \begin{minipage}[b]{0.40\textwidth}
        \centering
        \includegraphics[width=\textwidth]{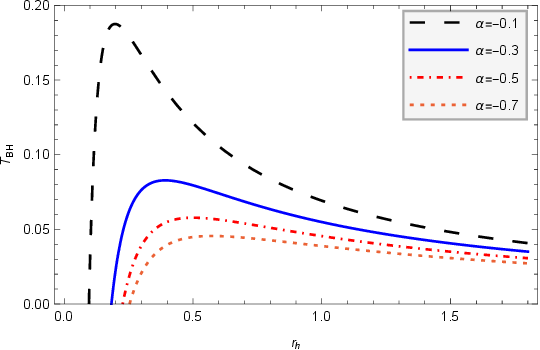}
    \end{minipage}
\caption{Equation (\ref{eqT1}) is plotted as a function of $r_{\rm{h}}$ with $D=1$ (left), $D=-1$ (right), $\omega=0$, and several values of $\alpha$.}
\label{dust2}
\end{figure}

Substituting $\omega=0$ into Eq. (\ref{eqT}), we get the Hawking temperature as
\begin{eqnarray}
	\label{eqT1}		
	T_{\rm{BH}}=\frac{2-3\alpha\left[1+Dr_{\rm{h}}^{-\frac{2}{2-3\alpha}}\right]}{4\pi r_{\rm{h}}(2-3\alpha)}.
\end{eqnarray}	
Eq. (\ref{eqT1}) provide a family of curves depending on the values of parameters $\alpha$ and $r_{\rm{h}}$. In the Fig. \ref{dust2}, we plot a set of curves of the Hawking temperature $T_{\rm{BH}}$ as a function of $r_{\rm{h}}$ by taking some values of $\alpha$. In the left figure, the temperature increases when the value of $r_{\rm{h}}$ increases for $0\leq\alpha<2/3$; the temperature first decreases, reaches the minimum value, and then starts to increase when $r_{\rm{h}}$ increases for $2/3<\alpha<1$. In the right figure, the temperature first increases, reaches the maximum value, and then starts to decrease when $r_{\rm{h}}$ increases for $-1<\alpha\leq 0$. The nonnegativity of temperature impose new limitations on the parameters of $\alpha$, $D$, and $r_{\rm h}$: $\alpha\left[1+Dr_{\rm{h}}^{-\frac{2}{2-3\alpha}}\right]\leq2/3$ for $-1<\alpha<2/3$ and $\alpha\left[1+Dr_{\rm{h}}^{-\frac{2}{2-3\alpha}}\right]\geq2/3$ for $2/3<\alpha<1$.

\subsection{Black hole surrounded by a Radiation field}
When BH surrounded by a radiation field, $\omega=1/3$, Eq. (\ref{eqB}) reduces to the following form
\begin{eqnarray}
	\label{eqB2}	
	B(r)=1-\frac{2M}{r}+Dr^{-1-\frac{6}{6-8\alpha}},
\end{eqnarray}
where $\alpha \neq 3/4$ to avoid the exponent singularity. Because the exist of $\alpha$, The metric linked to Eq. (\ref{eqB2}) differs from the metric of BH encompassed by the radiation field in GR \cite{Kiselev:2002dx}. In the limit of $\alpha\rightarrow 0$, we get the $B(r)=1-\frac{2M}{r}+\frac{D}{r^{2}}$, which is the effective metric of Reissner–Nordstr\"{o}m BH with an effective charge $Q^{2}_{\rm{eff}}=D$. %

If $\omega=1/3$, The conditions for SEC from Eqs. (\ref{sec1}) and (\ref{sec2}) take the following form
\begin{eqnarray}
	\label{sec021}	
	\frac{8(1-\alpha)D}{(6-8\alpha)^{2}}\geq0,\\
	\label{sec022}
	\frac{12(1-\alpha)D}{(6-8\alpha)^{2}}\geq0.
\end{eqnarray}	
Thinking of $|\alpha|<1$, these equations imply $\alpha\neq3/4$ for $D\geq0$. And there are no solutions for $D<0$.
\begin{figure}[H]
\centering
\includegraphics[width=10cm]{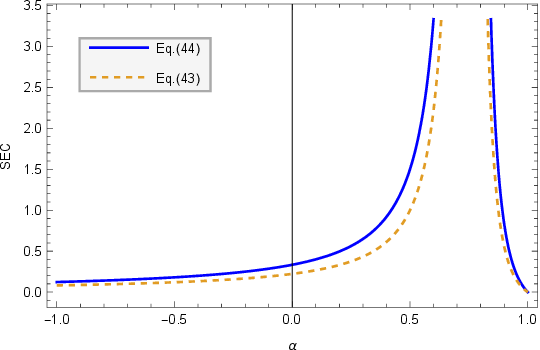}
\caption{Curves given by Eqs. (\ref{sec021}) and (\ref{sec022}) are plotted as a function of $\alpha$ with $D=1$ and $\omega=1/3$.}
\label{radi1}
\end{figure}
\begin{figure}[H]
	\centering
	\includegraphics[width=10cm]{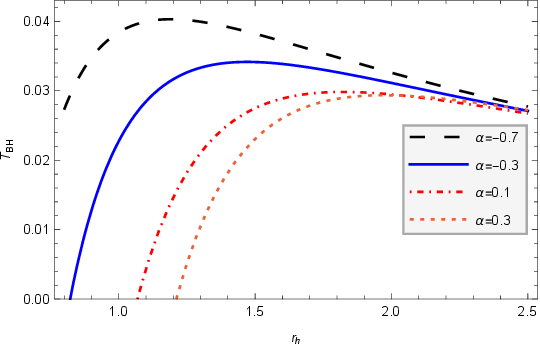}
	\caption{Equation (\ref{eqT2}) is plotted as a function of $r_{\rm{h}}$ with $D=1$, $\omega= 1/3$, and some values of $\alpha$ restricted}
	\label{radi2}
\end{figure}
 The regions where the SEC hold true are plotted with $D=1$ and $\omega=1/3$ in Fig. \ref{radi1}. The SEC is satisfied everywhere except at $\alpha=3/4$ where there is a divergence in the equations for the SEC. Obviously, the SEC is violated everywhere for $D<0$.

Taking $\omega=1/3$ in Eq. (\ref{eqT}), we get the Hawking temperature for the BH surrounded by a radiation field as
\begin{eqnarray}
	\label{eqT2}		
		T_{\rm{BH}}=\frac{1}{4\pi r_{\rm{h}}}-\frac{3Dr_{\rm{h}}^{-2-\frac{3}{3-4\alpha}}}{4\pi(3-4\alpha)}.
\end{eqnarray}	
In Fig. \ref{radi2}, we draw a family of curves of the Hawking temperature $T_{\rm{BH}}$ with respect to $r_{\rm{h}}$ for different values of $\alpha$. For all values of $\alpha$, the temperature $T_{\rm{BH}}$ should stay positive which gives additional limitations on the parameters of $\alpha$, $D$, and $r_{\rm h}$: $4\alpha+3Dr_{\rm{h}}^{-2-\frac{3}{3-4\alpha}}\leq3$ for $-1<\alpha<3/4$ and $4\alpha+3Dr_{\rm{h}}^{-2-\frac{3}{3-4\alpha}}\geq3$ for $3/4<\alpha<1$.

\subsection{Black hole surrounded by a quintessence field}

When BH surrounded by a quintessence field with $\omega=-2/3$, Eq. (\ref{eqB}) reduces to following form
\begin{eqnarray}
	\label{eqB3}	
	B(r)=1-\frac{2M}{r}+Dr^{\frac{2(3-8\alpha)}{6-11\alpha}},
\end{eqnarray}
where $\alpha\neq 6/11$ to avoid the exponent divergence. In the context of QFMG, we observe that there exists a family of solutions associated to this case.

\begin{figure}[H]
	\centering
	\includegraphics[width=10cm]{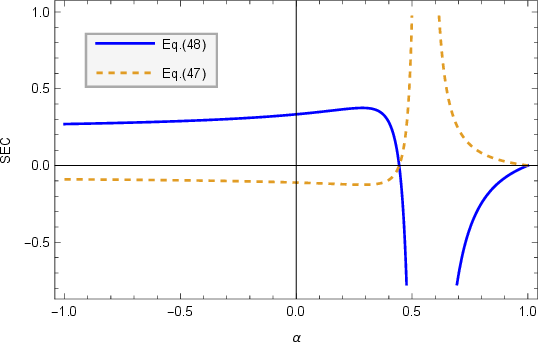}
	\caption{Conditions given by Eqs. (\ref{sec031}) and (\ref{sec032}) are plotted as a function of $\alpha$ with $D=1$.}
	\label{quin1}
\end{figure}

The conditions of SEC hold true for this case, Eqs. (\ref{sec1}) and (\ref{sec2}), take the form
\begin{eqnarray}
	\label{sec031}	
	\frac{(9\alpha-4)(1-\alpha)D}{(6-11\alpha)^{2}}\geq0,
\end{eqnarray}	
\begin{eqnarray}
	\label{sec032}	
	\frac{-3(9\alpha-4)(1-\alpha)D}{(6-11\alpha)^{2}}\geq0.
\end{eqnarray}	
Obviously, these expressions depend on the signal of the constant $D$.
The LHS of Eq. (\ref{sec032}) is $-3$ times to the LHS of Eq. (\ref{sec031}), namely, the conditions for the SEC holds true are given by these two equations that have different behavior in the domain under consideration. We observe that $\alpha\neq6/11$ and there is only one point given by $\alpha=4/9$ where Eqs. (\ref{sec031}) and (\ref{sec032}) hold true when $D$ takes positive values, as shown in Fig. \ref{quin1}. If $D$ takes negative values, the shapes of curves are reversed.

\begin{figure}[H]
	\centering
	\includegraphics[width=10cm]{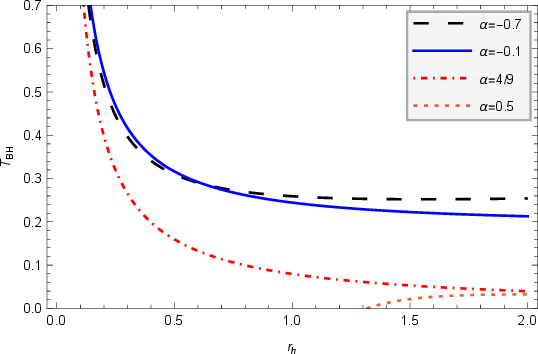}
	\caption{Equation (\ref{eqT3}) is plotted as a function of $r_{\rm{h}}$ with $D=1$ and $\omega=-2/3$.}
	\label{quin2}
\end{figure}

Taking $\omega=-2/3$ in Eq. (\ref{eqT}), the Hawking temperature for the BH surrounded by a quintessence field has the form
\begin{eqnarray}
	\label{eqT3}		
	T_{\rm{BH}}=\frac{6-11\alpha-3D(9\alpha-4)r_{\rm{h}}^{\frac{2(3-8\alpha)}{6-11\alpha}}}{4\pi r_{\rm{h}}(6-11\alpha)}.
\end{eqnarray}	
In Fig. \ref{quin2}, we plot the Hawking temperature $T_{\rm{BH}}$ as a function of $r_{\rm{h}}$ for some values
of $\alpha$. We observe that the shapes of curves change for each value of $\alpha$. We note, however, only on the curve with $\alpha=4/9$ the SEC holds true and the temperature is positive.

\subsection{Black hole surrounded by a Cosmological constant field}
When the BH surrounded by a cosmological constant field, $\omega=-1$, the Eq. (\ref{eqB}) reduces to the form
\begin{eqnarray}
	\label{eqB4}	
	B(r)=1-\frac{2M}{r}+Dr^{2},
\end{eqnarray}
which is a Schwarzschild-(anti) de Sitter BH. This solution is independent of the parameter $\alpha$ and is the same obtained in GR \cite{Kiselev:2002dx} and in $f(R,T)$ theory \cite{Santos:2023fgd}. However, the SEC in these three cases are all different.

The conditions for SEC hold represented by Eqs. (\ref{sec1}) and (\ref{sec2}) in this case are given by
\begin{eqnarray}
	\label{sec04}	
	\frac{(1-\alpha)D}{1-2\alpha}\geq 0,
\end{eqnarray}	
with Eq. (\ref{sec2}) being zero, which results in $-1<\alpha<1/2$ for $D>0$ and $1/2<\alpha<1$ for $D<0$.
\begin{figure}[H]
	\centering
	\includegraphics[width=10cm]{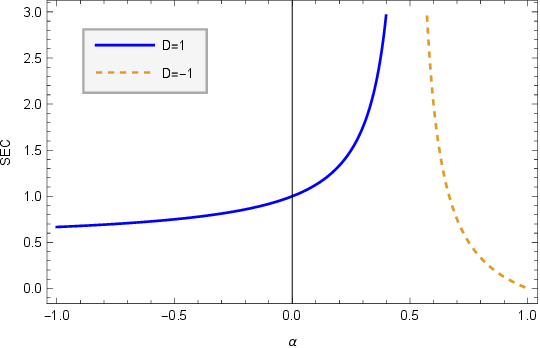}
	\caption{Condition (\ref{sec04}) is plotted as a function of $\alpha$ with $D=1$ or $D=-1$.}
	\label{cosm1}
\end{figure}

\begin{figure}[H]
	\centering
	\includegraphics[width=10cm]{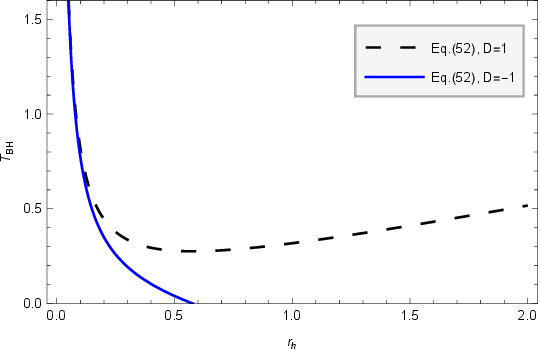}
	\caption{Equation (\ref{eqT4}) is plotted as a function of $r_{\rm{h}}$ with $D=1$ and $\omega=-1$.}
	\label{cosm2}
\end{figure}

Taking $\omega=-1$ in Eq. (\ref{eqT}), the Hawking temperature for the BH surrounded by a cosmological constant field recduces to
\begin{eqnarray}
	\label{eqT4}		
	T_{\rm{BH}}=\frac{1+3Dr_{\rm{h}}^{2}}{4\pi r_{\rm{h}}},
	\end{eqnarray}	
which is also independent of the parameter $\alpha$. In Fig. \ref{cosm2}, we plot the Hawking temperature as a function of $r_{\rm{h}}$. We observe that as $r_{\rm{h}}$ increases, the temperature firstly decreases, reaches its minimum value, and then begins to increase for the case of $D=1$. For $D=-1$, the temperature continuously decreases. The nonnegativity of temperature gives new restrictions on the parameters: $Dr^{2}_{\rm{h}}\geq -1/3$.

\subsection{Black hole surrounded by a Phantom field}
When the BH surrounded by a phantom field with $\omega=-4/3$, Eq. (\ref{eqB}) takes the form
\begin{eqnarray}
	\label{eqB5}	
	B(r)=1-\frac{2M}{r}+Dr^{\frac{2(9-16\alpha)}{6-13\alpha}},
\end{eqnarray}
which is dependent on the parameter $\alpha$ ($\alpha\neq 6/13$ to avoid the exponent singularity), implying that this solution in the context of QFMG is not equivalent to the solution related to $\omega=-4/3$ in GR \cite{Kiselev:2002dx}. For $\omega=-4/3$, the conditions for SEC hold are given by the relations
\begin{eqnarray}
	\label{sec051}	
	\frac{(8-15\alpha)(1-\alpha)D}{(6-13\alpha)^{2}}\geq0,
\end{eqnarray}	
\begin{eqnarray}
	\label{sec052}	
	\frac{9(8-15\alpha)(1-\alpha)D}{(6-13\alpha)^{2}}\geq0.
\end{eqnarray}

This pair of equations above demands that $-1<\alpha\leq8/15$ and $\alpha\neq6/13$ for positive values of $D$ and $8/15\leq\alpha<1$ for negative values of $D$.

\begin{figure}[H]
	\centering
	\includegraphics[width=10cm]{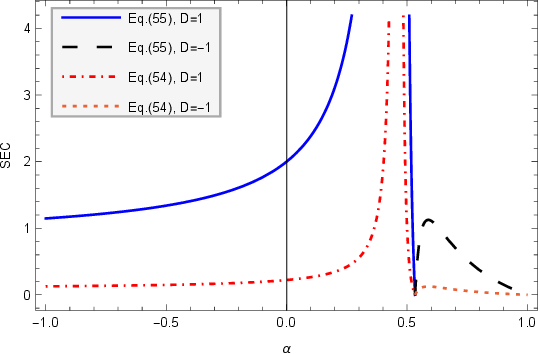}
	\caption{Conditions (\ref{sec051}) and (\ref{sec052}) are plotted as a function of $\alpha$ with $D=1$ or $D=-1$. }
	\label{phan1}
\end{figure}

We plot the LHSs of Eqs. (\ref{sec051}) and (\ref{sec052}) in Fig. \ref{phan1}, represented by the solid and dashed lines, respectively. We observe that they have similar behavior for $-1<\alpha\leq8/15$ and behave differently for $8/15\leq\alpha<1$. Positive values of the expressions are associated to the regions where the SEC holds true.

\begin{figure}[h]
    \centering
    \begin{minipage}[b]{0.40\textwidth}
        \centering
        \includegraphics[width=\textwidth]{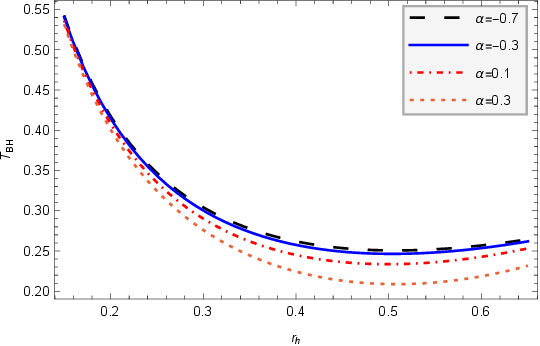}
    \end{minipage}
     \hspace{3em}
    \begin{minipage}[b]{0.40\textwidth}
        \centering
        \includegraphics[width=\textwidth]{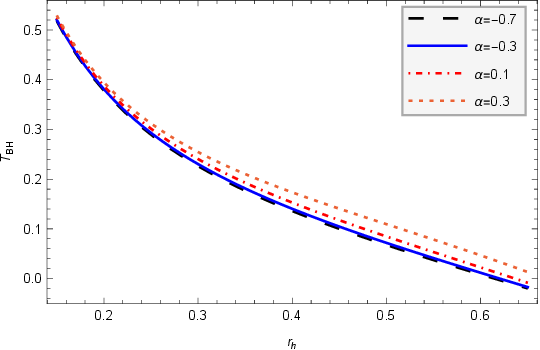}
    \end{minipage}
\caption{The Hawking temperature $T_{\rm{BH}}$, Eq. (\ref{eqT5}), is plotted as a function of $r_{\rm{h}}$ with $D=1$ (left), $D=-1$ (right),  $\omega=-4/3$, and some values of $\alpha$.}
\label{phan2}
\end{figure}

Taking $\omega=-4/3$ in Eq. (\ref{eqT}), the Hawking temperature $T_{\rm{BH}}$ for the BH surrounded
by a phantom field takes the form as follow
\begin{eqnarray}
	\label{eqT5}		
	T_{\rm{BH}}=\frac{6-13\alpha-3D(15\alpha-8)r_{\rm{h}}^{\frac{2(9-16\alpha)}{6-13\alpha}}}{4\pi r_{\rm{h}}(6-13\alpha)}.
\end{eqnarray}	
In this case, We plot the Hawking temperature $T_{\rm{BH}}$ in Fig. \ref{phan2}. The curves for each value of $\alpha$ have the same trend and little differences. In the domain studied, all the curves are associated to the positive values of temperature. The nonnegativity of temperature gives additional constraints on the parameters of of $\alpha$, $D$, and $r_{\rm D}$: $13\alpha+3D(15\alpha-8)r_{\rm{h}}^{\frac{2(9-16\alpha)}{6-13\alpha}}\leq6$ for $-1<\alpha<6/13$ and $r_{\rm h}$: $13\alpha+3D(15\alpha-8)r_{\rm{h}}^{\frac{2(9-16\alpha)}{6-13\alpha}}\geq6$ for $6/13<\alpha<1$.

\section{Conclusions and discussions}
We have considered the solution to the gravitational field equations in QFMG corresponding to the fluid of Kiselev’s. The general solution obtained has an additional structure that comes from the
dependence on the parameter of QFMG. This property implies that several
particular values of the parameter of EoS in QFMG lead to solutions which differs from the Kiselev black hole in GR.

We have also discussed the conditions that the fluid must satisfy in order to generate Kiselev black holes in the context of QFMG. For particular solutions corresponding to black holes surrounded by dust field, radiation field, quintessence field, cosmological constant field, and phantom field, respectively. We have analyzed in details the conditions imposed by SEC on the parameters. We have studied the horizons associated to these solutions and determined the Hawking temperatures. We also have discussed the limitations imposed by the positivity of temperature on the parameters.

Future researches can focus on discussing the influence of the parameter characterizing the quantum fluctuations of metric on the particle's motion, the accretion, the shadowing of BH, and so on.

\begin{acknowledgments}
This study is supported in part by National Natural Science Foundation of China (Grant No. 12333008).\\

\textbf{Data Availability Statement}: No Data associated in the manuscript.

\end{acknowledgments}
\bibliographystyle{elsarticle-num}
\bibliography{refq}

\begin{thebibliography}{10}
\expandafter\ifx\csname url\endcsname\relax
  \def\url#1{\texttt{#1}}\fi
\expandafter\ifx\csname urlprefix\endcsname\relax\def\urlprefix{URL }\fi
\expandafter\ifx\csname href\endcsname\relax
  \def\href#1#2{#2} \def\path#1{#1}\fi

\bibitem{Dzhunushaliev:2013nea}
V.~Dzhunushaliev, V.~Folomeev, B.~Kleihaus, J.~Kunz, {Modified gravity from the
  quantum part of the metric}, Eur.Phys.J. C74 (2014) 2743.
\newblock \href {http://arxiv.org/abs/1312.0225} {\path{arXiv:1312.0225}},
  \href {http://dx.doi.org/10.1140/epjc/s10052-014-2743-4}
  {\path{doi:10.1140/epjc/s10052-014-2743-4}}.

\bibitem{Dzhunushaliev:2015mva}
V.~Dzhunushaliev, V.~Folomeev, B.~Kleihaus, J.~Kunz, {Modified gravity from the
  nonperturbative quantization of a metric}, Eur.Phys.J. C75~(4) (2015) 157.
\newblock \href {http://arxiv.org/abs/1501.00886} {\path{arXiv:1501.00886}},
  \href {http://dx.doi.org/10.1140/epjc/s10052-015-3398-5}
  {\path{doi:10.1140/epjc/s10052-015-3398-5}}.

\bibitem{Dzhunushaliev:2015hoa}
V.~Dzhunushaliev, {Nonperturbative quantization: ideas, perspectives, and
  applications}\href {http://arxiv.org/abs/1505.02747}
  {\path{arXiv:1505.02747}}.

\bibitem{Yang:2015jla}
R.~Yang, {Effects of quantum fluctuations of metric on the universe}, Phys.
  Dark Univ. 13 (2016) 87--91.
\newblock \href {http://arxiv.org/abs/1506.02889} {\path{arXiv:1506.02889}},
  \href {http://dx.doi.org/10.1016/j.dark.2016.04.007}
  {\path{doi:10.1016/j.dark.2016.04.007}}.

\bibitem{Liu:2016qfx}
X.~Liu, T.~Harko, S.-D. Liang, {Cosmological implications of modified gravity
  induced by quantum metric fluctuations}, Eur. Phys. J. C 76~(8) (2016) 420.
\newblock \href {http://arxiv.org/abs/1607.04874} {\path{arXiv:1607.04874}},
  \href {http://dx.doi.org/10.1140/epjc/s10052-016-4275-6}
  {\path{doi:10.1140/epjc/s10052-016-4275-6}}.

\bibitem{Bernardo:2021ynf}
R.~C. Bernardo, C.-Y. Chen, J.~Said~Levi, Y.-H. Kung, {Confronting
  quantum-corrected teleparallel cosmology with observations}, JCAP 04~(04)
  (2022) 052.
\newblock \href {http://arxiv.org/abs/2111.11761} {\path{arXiv:2111.11761}},
  \href {http://dx.doi.org/10.1088/1475-7516/2022/04/052}
  {\path{doi:10.1088/1475-7516/2022/04/052}}.

\bibitem{Chen:2021oal}
C.-Y. Chen, Y.-H. Kung, {Modified Teleparallel Gravity induced by quantum
  fluctuations}, Phys. Dark Univ. 35 (2022) 100956.
\newblock \href {http://arxiv.org/abs/2108.04853} {\path{arXiv:2108.04853}},
  \href {http://dx.doi.org/10.1016/j.dark.2022.100956}
  {\path{doi:10.1016/j.dark.2022.100956}}.

\bibitem{Lima:2023qdv}
F.~C.~E. Lima, C.~A.~S. Almeida, {Effects of quantum fluctuations of the metric
  on a braneworld}\href {http://arxiv.org/abs/2307.05879}
  {\path{arXiv:2307.05879}}.

\bibitem{Yang:2020lxv}
J.-Z. Yang, S.~Shahidi, T.~Harko, S.-D. Liang, {Black hole solutions in
  modified gravity induced by quantum metric fluctuations}, Phys. Dark Univ. 31
  (2021) 100756.
\newblock \href {http://arxiv.org/abs/2012.02723} {\path{arXiv:2012.02723}},
  \href {http://dx.doi.org/10.1016/j.dark.2020.100756}
  {\path{doi:10.1016/j.dark.2020.100756}}.

\bibitem{Haghani:2021iqe}
Z.~Haghani, T.~Harko, {Effects of Quantum Metric Fluctuations on the
  Cosmological Evolution in Friedmann-Lemaitre-Robertson-Walker Geometries},
  MDPI Physics 3~(3) (2021) 689--714.
\newblock \href {http://dx.doi.org/10.3390/physics3030042}
  {\path{doi:10.3390/physics3030042}}.

\bibitem{Yang:2024gnf}
R.-J. Yang, Y.-B. Shi, {Baryogenesis in quantum fluctuation modified gravity},
  Phys. Dark Univ. 46 (2024) 101645.
\newblock \href {http://arxiv.org/abs/2402.07205} {\path{arXiv:2402.07205}},
  \href {http://dx.doi.org/10.1016/j.dark.2024.101645}
  {\path{doi:10.1016/j.dark.2024.101645}}.

\bibitem{Kiselev:2002dx}
V.~V. Kiselev, {Quintessence and black holes}, Class. Quant. Grav. 20 (2003)
  1187--1198.
\newblock \href {http://arxiv.org/abs/gr-qc/0210040}
  {\path{arXiv:gr-qc/0210040}}, \href
  {http://dx.doi.org/10.1088/0264-9381/20/6/310}
  {\path{doi:10.1088/0264-9381/20/6/310}}.

\bibitem{Santos:2023fgd}
L.~C.~N. Santos, F.~M. da~Silva, C.~E. Mota, I.~P. Lobo, V.~B. Bezerra,
  {Kiselev black holes in f(R,~T) gravity}, Gen. Rel. Grav. 55~(8) (2023) 94.
\newblock \href {http://arxiv.org/abs/2301.02534} {\path{arXiv:2301.02534}},
  \href {http://dx.doi.org/10.1007/s10714-023-03138-z}
  {\path{doi:10.1007/s10714-023-03138-z}}.

\bibitem{Ghosh:2015ovj}
S.~G. Ghosh, {Rotating black hole and quintessence}, Eur. Phys. J. C 76~(4)
  (2016) 222.
\newblock \href {http://arxiv.org/abs/1512.05476} {\path{arXiv:1512.05476}},
  \href {http://dx.doi.org/10.1140/epjc/s10052-016-4051-7}
  {\path{doi:10.1140/epjc/s10052-016-4051-7}}.

\bibitem{Xu:2016jod}
Z.~Xu, J.~Wang, {Kerr-Newman-AdS Black Hole In Quintessential Dark Energy},
  Phys. Rev. D 95~(6) (2017) 064015.
\newblock \href {http://arxiv.org/abs/1609.02045} {\path{arXiv:1609.02045}},
  \href {http://dx.doi.org/10.1103/PhysRevD.95.064015}
  {\path{doi:10.1103/PhysRevD.95.064015}}.

\bibitem{Toshmatov:2015npp}
B.~Toshmatov, Z.~Stuchl\'\i{}k, B.~Ahmedov, {Rotating black hole solutions with
  quintessential energy}, Eur. Phys. J. Plus 132~(2) (2017) 98.
\newblock \href {http://arxiv.org/abs/1512.01498} {\path{arXiv:1512.01498}},
  \href {http://dx.doi.org/10.1140/epjp/i2017-11373-4}
  {\path{doi:10.1140/epjp/i2017-11373-4}}.

\bibitem{Ghosh:2023nkr}
S.~G. Ghosh, S.~U. Islam, S.~D. Maharaj, {Rotating kiselev black holes in f(R,
  T) gravity}, Phys. Scripta 99~(6) (2024) 065032.
\newblock \href {http://arxiv.org/abs/2307.11611} {\path{arXiv:2307.11611}},
  \href {http://dx.doi.org/10.1088/1402-4896/ad4833}
  {\path{doi:10.1088/1402-4896/ad4833}}.

\bibitem{Singh:2023zmy}
B.~P. Singh, {Shadows of quintessential dark energy black holes in the domain
  of outer communication}, Phys. Dark Univ. 42 (2023) 101279.
\newblock \href {http://arxiv.org/abs/2301.00956} {\path{arXiv:2301.00956}},
  \href {http://dx.doi.org/10.1016/j.dark.2023.101279}
  {\path{doi:10.1016/j.dark.2023.101279}}.

\bibitem{Hamil:2024ppj}
B.~Hamil, B.~C. L\"utf\"uo\u{g}lu, {Noncommutative Schwarzschild black hole
  surrounded by quintessence: Thermodynamics, Shadows and Quasinormal modes},
  Phys. Dark Univ. 44 (2024) 101484.
\newblock \href {http://arxiv.org/abs/2401.09295} {\path{arXiv:2401.09295}},
  \href {http://dx.doi.org/10.1016/j.dark.2024.101484}
  {\path{doi:10.1016/j.dark.2024.101484}}.

\bibitem{Giambo:2023zmy}
R.~Giamb\`o, O.~Luongo, {De Sitter-like configurations with asymptotic
  quintessence environment}, Class. Quant. Grav. 41~(12) (2024) 125005.
\newblock \href {http://arxiv.org/abs/2308.10060} {\path{arXiv:2308.10060}},
  \href {http://dx.doi.org/10.1088/1361-6382/ad43a9}
  {\path{doi:10.1088/1361-6382/ad43a9}}.

\bibitem{Qu:2023rsv}
Z.-S. Qu, T.~Wang, C.-J. Feng, {Reduced Kiselev black hole}, Eur. Phys. J. C
  83~(9) (2023) 784.
\newblock \href {http://arxiv.org/abs/2307.09079} {\path{arXiv:2307.09079}},
  \href {http://dx.doi.org/10.1140/epjc/s10052-023-11962-5}
  {\path{doi:10.1140/epjc/s10052-023-11962-5}}.

\bibitem{Heydarzade:2023dof}
Y.~Heydarzade, M.~Misyura, V.~Vertogradov, {Hairy Kiselev black hole
  solutions}, Phys. Rev. D 108~(4) (2023) 044073.
\newblock \href {http://arxiv.org/abs/2307.04556} {\path{arXiv:2307.04556}},
  \href {http://dx.doi.org/10.1103/PhysRevD.108.044073}
  {\path{doi:10.1103/PhysRevD.108.044073}}.

\bibitem{Benali:2024clh}
M.~Benali, A.~E. Balali, {Rotating reduced Kiselev black holes: Shadows, Energy
  emission and Deflection of light}\href {http://arxiv.org/abs/2406.00788}
  {\path{arXiv:2406.00788}}.

\bibitem{Sood:2024ufi}
A.~Sood, A.~Kumar, J.~K. Singh, S.~G. Ghosh, {Photon orbits and phase
  transitions in Kiselev-AdS black holes from $f(R,\; T)$ gravity}, Eur. Phys.
  J. C 84~(8) (2024) 876.
\newblock \href {http://dx.doi.org/10.1140/epjc/s10052-024-13251-1}
  {\path{doi:10.1140/epjc/s10052-024-13251-1}}.

\bibitem{Visser:2019brz}
M.~Visser, {The Kiselev black hole is neither perfect fluid, nor is it
  quintessence}, Class. Quant. Grav. 37~(4) (2020) 045001.
\newblock \href {http://arxiv.org/abs/1908.11058} {\path{arXiv:1908.11058}},
  \href {http://dx.doi.org/10.1088/1361-6382/ab60b8}
  {\path{doi:10.1088/1361-6382/ab60b8}}.

\bibitem{Bezerra:2022srj}
V.~B. Bezerra, L.~C.~N. Santos, F.~M. da~Silva, H.~Moradpour, {On black holes
  surrounded by a fluid of strings in Rastall gravity}, Gen. Rel. Grav. 54~(9)
  (2022) 109.
\newblock \href {http://arxiv.org/abs/2405.20966} {\path{arXiv:2405.20966}},
  \href {http://dx.doi.org/10.1007/s10714-022-02993-6}
  {\path{doi:10.1007/s10714-022-02993-6}}.

\bibitem{Lungu:2024iob}
V.~Lungu, M.-A. Dariescu, {Charged Particles Orbiting a Weakly Magnetized Black
  Hole Immersed in Quintessential Matter}\href
  {http://arxiv.org/abs/2405.14420} {\path{arXiv:2405.14420}}.

\bibitem{Chaudhary:2024yod}
S.~Chaudhary, M.~D. Sultan, A.~Malik, A.~ur~Rehman, A.~\"Ovg\"un, A.~A. Ghfar,
  {Images and stability of black hole with cloud of strings and quintessence in
  EGUP framework}, Nucl. Phys. B 1006 (2024) 116635.
\newblock \href {http://dx.doi.org/10.1016/j.nuclphysb.2024.116635}
  {\path{doi:10.1016/j.nuclphysb.2024.116635}}.

\bibitem{Gogoi:2024ypn}
D.~J. Gogoi, Y.~Sekhmani, S.~Bora, J.~Rayimbaev, J.~Bora, R.~Myrzakulov,
  {Corrected thermodynamics and stability of magnetic charged AdS black holes
  surrounded by quintessence}, JCAP 11 (2024) 019.
\newblock \href {http://arxiv.org/abs/2407.10946} {\path{arXiv:2407.10946}},
  \href {http://dx.doi.org/10.1088/1475-7516/2024/11/019}
  {\path{doi:10.1088/1475-7516/2024/11/019}}.

\bibitem{Hamil:2022pzr}
B.~Hamil, B.~C. L\"utf\"uo\u{g}lu, {Thermodynamics of Schwarzschild black hole
  surrounded by quintessence in gravity's rainbow}, Nucl. Phys. B 990 (2023)
  116191.
\newblock \href {http://arxiv.org/abs/2209.00960} {\path{arXiv:2209.00960}},
  \href {http://dx.doi.org/10.1016/j.nuclphysb.2023.116191}
  {\path{doi:10.1016/j.nuclphysb.2023.116191}}.

\bibitem{Sheikhahmadi:2023jpb}
H.~Sheikhahmadi, S.~Soroushfar, S.~N. Sajadi, T.~Harko, {Astrophysical and
  electromagnetic emissivity properties of black holes surrounded by a
  quintessence type exotic fluid in the
  scalar\textendash{}vector\textendash{}tensor modified gravity}, Eur. Phys. J.
  C 83~(9) (2023) 814.
\newblock \href {http://arxiv.org/abs/2303.02194} {\path{arXiv:2303.02194}},
  \href {http://dx.doi.org/10.1140/epjc/s10052-023-11980-3}
  {\path{doi:10.1140/epjc/s10052-023-11980-3}}.

\bibitem{Zahid:2023csk}
M.~Zahid, J.~Rayimbaev, F.~Sarikulov, S.~U. Khan, J.~Ren, {Shadow of rotating
  and twisting charged black holes with cloud of strings and quintessence},
  Eur. Phys. J. C 83~(9) (2023) 855.
\newblock \href {http://dx.doi.org/10.1140/epjc/s10052-023-12025-5}
  {\path{doi:10.1140/epjc/s10052-023-12025-5}}.

\bibitem{Hamil:2024njs}
B.~Hamil, B.~C. L\"utf\"uo\u{g}lu, {Euler-Heisenberg black hole surrounded by
  quintessence in the background of perfect fluid dark matter: Thermodynamics,
  Shadows and Quasinormal modes}\href {http://arxiv.org/abs/2406.02109}
  {\path{arXiv:2406.02109}}.

\bibitem{Ghosh:2017cuq}
S.~G. Ghosh, S.~D. Maharaj, D.~Baboolal, T.-H. Lee, {Lovelock black holes
  surrounded by quintessence}, Eur. Phys. J. C 78~(2) (2018) 90.
\newblock \href {http://arxiv.org/abs/1708.03884} {\path{arXiv:1708.03884}},
  \href {http://dx.doi.org/10.1140/epjc/s10052-018-5570-1}
  {\path{doi:10.1140/epjc/s10052-018-5570-1}}.

\bibitem{Heydarzade:2017wxu}
Y.~Heydarzade, F.~Darabi, {Black Hole Solutions Surrounded by Perfect Fluid in
  Rastall Theory}, Phys. Lett. B 771 (2017) 365--373.
\newblock \href {http://arxiv.org/abs/1702.07766} {\path{arXiv:1702.07766}},
  \href {http://dx.doi.org/10.1016/j.physletb.2017.05.064}
  {\path{doi:10.1016/j.physletb.2017.05.064}}.

\bibitem{Sakti:2019krw}
M.~F. A.~R. Sakti, A.~Suroso, F.~P. Zen,
  {Kerr\textendash{}Newman\textendash{}NUT\textendash{}Kiselev black holes in
  Rastall theory of gravity and Kerr/CFT correspondence}, Annals Phys. 413
  (2020) 168062.
\newblock \href {http://arxiv.org/abs/1901.09163} {\path{arXiv:1901.09163}},
  \href {http://dx.doi.org/10.1016/j.aop.2019.168062}
  {\path{doi:10.1016/j.aop.2019.168062}}.

\bibitem{Wetterich:2016vxu}
C.~Wetterich, {Quantum correlations for the metric}, Phys. Rev. D 95~(12)
  (2017) 123525.
\newblock \href {http://arxiv.org/abs/1603.06504} {\path{arXiv:1603.06504}},
  \href {http://dx.doi.org/10.1103/PhysRevD.95.123525}
  {\path{doi:10.1103/PhysRevD.95.123525}}.

\bibitem{Martin-Moruno:2017exc}
P.~Martin-Moruno, M.~Visser, {Classical and semi-classical energy conditions},
  Fundam. Theor. Phys. 189 (2017) 193--213.
\newblock \href {http://arxiv.org/abs/1702.05915} {\path{arXiv:1702.05915}},
  \href {http://dx.doi.org/10.1007/978-3-319-55182-1_9}
  {\path{doi:10.1007/978-3-319-55182-1_9}}.

\bibitem{Barcelo:2002bv}
C.~Barcelo, M.~Visser, {Twilight for the energy conditions?}, Int. J. Mod.
  Phys. D 11 (2002) 1553--1560.
\newblock \href {http://arxiv.org/abs/gr-qc/0205066}
  {\path{arXiv:gr-qc/0205066}}, \href
  {http://dx.doi.org/10.1142/S0218271802002888}
  {\path{doi:10.1142/S0218271802002888}}.

\bibitem{Martin-Moruno:2013wfa}
P.~Martin-Moruno, M.~Visser, {Semiclassical energy conditions for quantum
  vacuum states}, JHEP 09 (2013) 050.
\newblock \href {http://arxiv.org/abs/1306.2076} {\path{arXiv:1306.2076}},
  \href {http://dx.doi.org/10.1007/JHEP09(2013)050}
  {\path{doi:10.1007/JHEP09(2013)050}}.

\bibitem{Martin-Moruno:2013sfa}
P.~Mart{\'\i}n-Moruno, M.~Visser, {Classical and quantum flux energy conditions
  for quantum vacuum states}, Phys. Rev. D 88~(6) (2013) 061701.
\newblock \href {http://arxiv.org/abs/1305.1993} {\path{arXiv:1305.1993}},
  \href {http://dx.doi.org/10.1103/PhysRevD.88.061701}
  {\path{doi:10.1103/PhysRevD.88.061701}}.

\bibitem{Visser:1992qh}
M.~Visser, {Dirty black holes: Thermodynamics and horizon structure}, Phys.
  Rev. D 46 (1992) 2445--2451.
\newblock \href {http://arxiv.org/abs/hep-th/9203057}
  {\path{arXiv:hep-th/9203057}}, \href
  {http://dx.doi.org/10.1103/PhysRevD.46.2445}
  {\path{doi:10.1103/PhysRevD.46.2445}}.

\bibitem{Vertogradov:2024fbd}
V.~Vertogradov, {Dynamical black holes: Apparent horizon versus energy
  conditions}, Phys. Lett. B 867 (2025) 139607.
\newblock \href {http://arxiv.org/abs/2410.10582} {\path{arXiv:2410.10582}},
  \href {http://dx.doi.org/10.1016/j.physletb.2025.139607}
  {\path{doi:10.1016/j.physletb.2025.139607}}.

\end{thebibliography}

\end{document}